\documentclass[usenatbib]{mn2e}
\usepackage{graphicx}
\usepackage{caption}
\usepackage{multicol}
\usepackage{color}
\usepackage{adjustbox}
\begin{document}

\title[Transmission spectroscopy of WASP-52b]{Transmission spectroscopy of the inflated exoplanet WASP-52b, and evidence for a bright region on the stellar surface}

\author[J. Kirk et al.]{J. Kirk$^1$\thanks{James.Kirk@warwick.ac.uk}, 
P. J. Wheatley$^1$\thanks{P.J.Wheatley@warwick.ac.uk}, 
T. Louden$^1$, 
S. P. Littlefair$^2$, 
C. M. Copperwheat$^3$,
\newauthor D. J. Armstrong$^{1,4}$, 
T. R. Marsh$^1$ 
and V. S. Dhillon$^{2,5}$ 
\\ 
$^1$Department of Physics, University of Warwick, Coventry, CV4 7AL, UK \\ 
$^2$Department of Physics and Astronomy, University of Sheffield, Sheffield, S3 7RH, UK \\
$^3$Astrophysics Research Institute, Liverpool John Moores University, Liverpool, L3 5RF, UK \\
$^4$ARC, School of Mathematics and Physics, Queen's University Belfast, University Road, Belfast BT7 1NN, UK \\
$^5$Instituto de Astrof\'{i}sica de Canarias, V\'{i}a L\'{a}ctea s/n, 38205, La Laguna, Spain
}

\maketitle

\begin{abstract}

We have measured the transmission spectrum of the extremely inflated hot Jupiter WASP-52b using simultaneous photometric observations in SDSS $u'$, $g'$ and a filter centred on the sodium doublet (NaI) with the ULTRACAM instrument mounted on the 4.2m William Herschel Telescope. We find that Rayleigh scattering is not the dominant source of opacity within the planetary atmosphere and find a transmission spectrum more consistent with wavelength-independent opacity such as from clouds. We detect an in-transit anomaly that we attribute to the presence of stellar activity and find that this feature can be more simply modelled as a bright region on the stellar surface akin to Solar faculae rather than spots. A spot model requires a significantly larger planet/star radius ratio than that found in previous studies. Our results highlight the precision that can be achieved by ground-based photometry with errors in the scaled planetary radii of less than one atmospheric scale height, comparable to HST observations. 

\end{abstract}

\section{Introduction}

Transmission spectroscopy using HST/STIS was used to detect the first spectral feature of any exoplanet, that of the narrow core of atomic sodium originating from the upper atmosphere of HD\,209458b \citep{Charbonneau2002}. Broad-band opacity sources were later detected in this planet's atmosphere in the form of the pressure broadened wings of NaI \citep{Sing2008_spectra} and a blueward slope in the transmission spectrum interpreted as Rayleigh scattering by H$_2$ \citep{Etangs2008_HD209}. This feature was also seen in HST data for HD\,189733b (\citealt{Pont2008}; \citealt{Sing2011}), however the absence of the broad wings of NaI suggested the presence of silicate condensates in the upper atmosphere of the planet \citep{Etangs2008_HD189}. Recent studies of hot Jupiters with HST have revealed a diverse array of atmospheres from clear to cloudy (e.g. \citealt{Sing2016}). These studies have included the detections of Rayleigh scattering by high altitude hazes also in WASP-12b \citep{Sing2013}, additional unknown optical absorbers in HAT-P-1b (\citealt{Nikolov2014}), a cloud deck masking the majority of the NaI absorption and near infrared water features in WASP-31b \citep{Sing2015} and a clear atmosphere consistent with solar and sub-solar metallicities in WASP-39b (\citealt{Sing2016}; \citealt{Fischer2016}).

Ground-based measurements of hot Jupiter transmission signals have also found success, beginning with the detection of the narrow component of NaI in HD\,189733b \citep{Redfield2008} and HD\,209458b (\citealt{Snellen2008}; \citealt{Langland-Shula2009}). More recent results have included further detections of sodium (e.g. \citealt{Wood2011}; \citealt{Zhou2012}), potassium (e.g. \citealt{Sing2011_XO2b}; \citealt{Wilson2015}), blueward scattering slopes (e.g. \citealt{Jordan2013}; \citealt{Stevenson2014}), and atmospheres dominated by clouds or hazes (e.g. \citealt{Gibson2013_wasp29}; \citealt{Mallonn2016}).

Although there has been success with ground-based spectroscopic observations they are more susceptible to systematics arising from, for example, differential slit losses (e.g. \citealt{Sing2012}). A simpler approach, which avoids this problem, is to use photometers to perform simultaneous broadband multi-wavelength measurements. These studies have included evidence for large blueward slopes (e.g. \citealt{Southworth2012}; \citealt{Southworth2015}; \citealt{Mancini2016}), enhanced absorption around the alkali metal lines (e.g. \citealt{Mancini2013}; \citealt{Bento2014}) and transmission spectra most consistent with clouds (e.g. \citealt{Mallonn2015_hat12}). By observing the planetary radius at carefully selected wavelengths we can probe for the existence of Rayleigh scattering and the broad absorption wings of the NaI doublet that are expected to be present in the cloud free atmospheres of hot Jupiters \citep{SeagerSasselov}.

Transmission spectroscopy is often focussed on inflated hot Jupiters; planets with very low density due to their large radii and relatively low masses. As a result of their low densities, low surface gravities and high temperatures these planets have large atmospheric scale heights, $H$, given by

\begin{equation}
\label{eq:scaleheight}
H = \frac{kT}{g\mu}
\end{equation}

\noindent
where $k$ is Boltzmann's constant, $T$ is the temperature of the planet, $g$ is the acceleration due to gravity and $\mu$ is the mean molecular mass. The outermost $\sim$ 5 scale heights may account for up to 10\% of the cross sectional area of a hot Jupiter, where different atmospheric species can affect the observed transmission spectrum \citep{Brown2001}. The relative size of the planet's scale height to the stellar disc governs the amplitude of the transmission signal.

WASP-52b \citep{Hebrard2013} is an extremely inflated hot Jupiter, with a mass of 0.46\,M$\mathrm{_{J}}$ and radius of 1.27\,R$\mathrm{_{J}}$ giving it a mean density of 0.299\,g/cm$^3$. It orbits its 0.87\,M$_{\odot}$ K2V host star with a period of 1.75 days. Due to the combination of the inflated planetary radius with the small radius of the host (0.79\,R$_{\odot}$), it shows a deep transit in the WASP photometry (2.7\,\%). From the table of system parameters (Table \ref{tab:wasp52}; \citealt{Hebrard2013}), and assuming a Jupiter mean molecular mass of 2.3 times the mass of a proton, the scale height of WASP-52b is calculated to be 731\,km. This makes WASP-52b an exceptional target for transmission spectroscopy as the difference in the transit depth corresponding to one atmospheric scale height is $4.4 \times 10^{-4}$, at least three times stronger than that of HD\,189733b.

In this paper we present multi-wavelength observations of WASP-52b taken using the high speed multi-band photometer ULTRACAM \citep{Dhillon2007}.

\section{Observations}

WASP-52b was observed on the night of the 7th September 2012 using the ULTRACAM \citep{Dhillon2007} instrument on the 4.2m William Herschel Telescope (WHT), La Palma. ULTRACAM is a high speed triple beam CCD photometer. Incoming light is split into three bandpasses, using two dichroics, and re-imaged onto three CCDs at a resolution of 0.3" per pixel, with a field of view of 5'. 

ULTRACAM is particularly useful for the ground-based application of transmission spectroscopy as it simultaneously takes measurements at three different wavelengths, enabling the transit depth to be measured as a function of wavelength (e.g. \citealt{Copperwheat2013}; \citealt{Bento2014}). The use of frame transfer CCDs allow for high frame rates (up to 300\,Hz) with little dead time (24\,ms), which is useful to avoid saturation when observing bright stars and enables many more sky flats to be taken. 

The observations were made using SDSS $u'$ ($\lambda_{central}$ = 3557\,\AA, FWHM = 599\,\AA) and $g'$ ($\lambda_{central}$ = 4825\,\AA, FWHM = 1379\,\AA) filters and a filter centred on the NaI doublet ($\lambda_{central}$ = 5912\,\AA, FWHM = 312\,\AA). These filters were selected to probe for Rayleigh scattering by observing the difference in transit depth between the $u'$ and $g'$ bands and to search for the broad wings of the sodium doublet with the NaI filter.

The observations were performed with moderate telescope defocussing ($\sim 3$ arcsec) in windowed mode with exposure times of 0.76 seconds in the red and green channels and a cadence of 0.79 seconds. Due to the reduced photon count in the blue channel multiples of 10 frames were averaged on-chip before readout, leading to a 7.9 second exposure time in this channel.

The moon was at 54\% illumination on the night of our observations and we analysed the data with an airmass varying from 1.41 $\rightarrow$ 1.06 $\rightarrow$ 2.00.

All the data were reduced using the ULTRACAM data reduction pipeline\footnote{http://deneb.astro.warwick.ac.uk/phsaap/software/ultracam/ \\html/index.html} with bias subtraction and flatfielding performed in the standard way. Aperture photometry was performed for all frames using a fixed aperture. Initially, many reductions were performed with a variety of aperture sizes and the signal-to-noise ratio was calculated for each using the ratio of aperture counts to aperture errors. The optimal aperture radius was found to be 18 pixels with a sky annulus of inner radius 23 pixels and outer radius 27 pixels.

Some of the observations of WASP-52 were taken through cloud, seen as dips in transmission in the raw light curves, which had to be removed before analysis. These were well defined, discrete events, with good quality data taken between the clouds. The cloudy data were removed using an iterative process. An array of running medians was calculated over a sliding box of 600 data points in the red and green arms and 60 points in the blue. This array of running medians was then subtracted from the raw data to flatten it. The median absolute deviation (MAD) of this subtracted array was calculated and sigma clipping performed to remove those data points lying at $\geq 6\sigma$ from the MAD. This process was then repeated with a smaller sliding box of 400 data points in the red and green and 40 in the blue with a final sigma cut at 5$\sigma$ from the MAD. The sigma clipping was performed on both the target and comparison star independently and only those frames that passed the sigma cut for both were kept, resulting in the removal of 20\,\% of the data. After the sigma clipping and binning of the red and green channels to the cadence of the blue channel, there were 2494 data points in each of the three light curves at a cadence of 7.9 seconds giving us excellent sampling even after the cleaning of the data.

Differential photometry was subsequently performed to remove the worst effects of telluric extinction using a comparison star with similar magnitude and colour to WASP-52 (BD+08 5023, 23:14:12.026 +08:50:56.28). This star has a V magnitude of 10.59 and B-V colour of 0.86, whilst WASP-52 has a V magnitude of 12.22 and B-V colour of 0.82. The comparison star was checked and found to be photometrically stable. Combinations of fainter stars in the field were tested as comparison stars but led to more scatter in the differential light curve than division by the single, bright comparison.

\section{Data Analysis}

\subsection{Light curve fitting with analytic model}

\begin{figure*}
\centering
\includegraphics[scale=0.45]{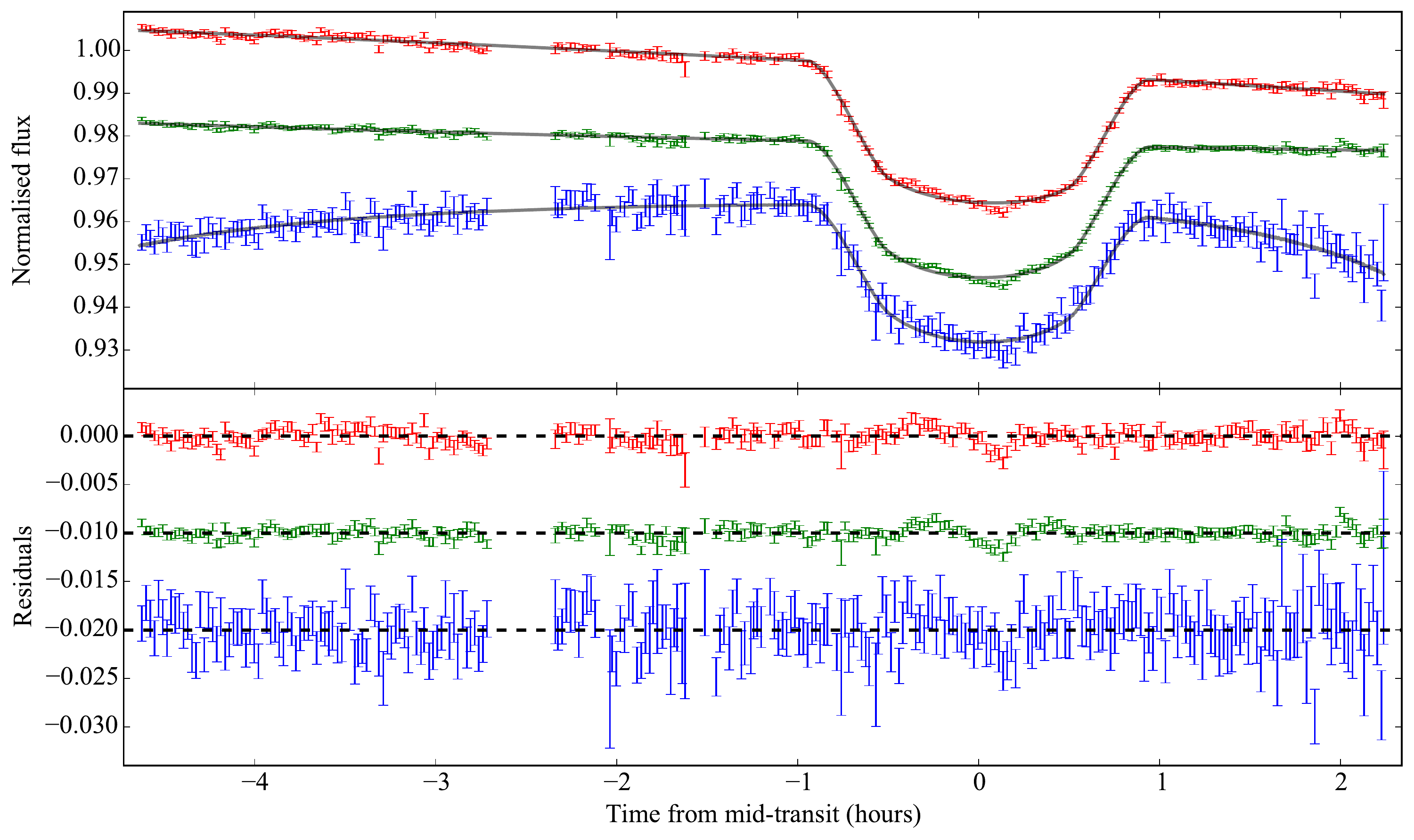}
\caption{MCMC fits of analytic quadratic limb-darkened transit light curves \protect\citep{MandelAgol} revealing the presence of an in-transit anomaly. The upper panel shows the fits to each of the 3 wavelengths; NaI (top), $g'$ (middle) and $u'$ (bottom) with the latter two light curves each offset by -0.02 for clarity. The lower panel shows the residuals from these fits with the latter two offset by -0.01 and -0.02, respectively.}
\label{fig:mandel_agol}
\end{figure*}

We initially fitted the differential light curves with analytic limb-darkened transit light curves \citep{MandelAgol} using a Markov Chain Monte Carlo (MCMC) algorithm, implemented through the \textsc{emcee} \citep{emcee} Python package. A quadratic limb darkening law was used and fit simultaneously with a long time-scale trend so as not to bias the derived radii. In the red and green channels, this trend was fit with a second order polynomial whilst in the blue it was fit as a function of airmass since it was clearly related to extinction. In order to fit the airmass term in the blue channel, an extinction coefficient that varied quadratically in time was used, which replicated the trend well. An extinction coefficient that varied linearly in time was also tested but could not fit the sharp downturn in the blue light curve at the end of the night (Fig. \ref{fig:mandel_agol}). 

The scaled semi major axis, $a/R_{*}$, inclination of the orbit and the time of mid transit, $T_{0}$, were tied across the three light curves when fitting. The parameters that were fit individually in each of the channels were the ratio of planet to star radius $R_{P}/R_{*}$, the second limb darkening coefficient $u2$, and the parameters defining the long time scale trend. The first limb darkening coefficient, $u1$, was held fixed in the fitting as there is a degeneracy between the limb darkening parameters which can affect the light curve solution \citep{Southworth2008}. The limb darkening coefficients and priors were chosen from the tables of \cite{Claret2011}. Uniform priors were adopted for all the model parameters, with the MCMC walkers started at the values from \cite{Hebrard2013}.

The resulting fits of the analytic limb-darkened transit light curves are shown in Fig. \ref{fig:mandel_agol}. The strongest residual across the whole light curve in all three wavelengths is seen during transit and is consistent across the three bands. This residual is akin to the planet occulting areas of stellar activity. We considered the possibility that this anomaly could have been associated with the use of incorrect limb darkening coefficients but no choice of coefficients could replicate this feature.

\subsection{Fitting of star spot model}
\label{sec:spot_model}

The presence of the in-transit anomaly after the fitting of limb-darkened analytic light curves motivated the use of spot models. Star spot occultations have been seen in the transit light curves of several planets, including HD\,189733b (\citealt{Pont2007}; \citealt{Sing2011}), TrES-1b (\citealt{Rabus2009}; \citealt{Dittmann2009}), CoRoT-2b (\citealt{Wolter2009}; \citealt{Huber2009}; \citealt{Silva-Valio2010}), HAT-P-11b (\citealt{Sanchis-Ojeda2011_hat11}; \citealt{Deming2011}; \citealt{Beky2014_hat11}), WASP-4b (\citealt{Sanchis-Ojeda2011_wasp4}; \citealt{Hoyer2013}), WASP-19b (\citealt{Mancini2013}; \citealt{Reed2013}; \citealt{Huitson2013}; \citealt{Mandell2013}; \citealt{Sedaghati2015}), HATS-2b \citep{Mohler2013}, Kepler-63b \citep{Sanchis-Ojeda2013}, Qatar-2b \citep{Mancini2014}, and HAT-P-36b \citep{Mancini2015}. 

If a planet occults a spot (a region cooler than the surrounding photosphere) it blocks less of the stellar flux than compared with its transit across the hotter pristine stellar disc. This results in a bump during transit and therefore a smaller derived planetary radius. Star spot activity is not unexpected for WASP-52 since \cite{Hebrard2013} found modulations in its light curve and chromospheric emission peaks in the CaII H+K lines. Using these modulations they calculated the rotation period of WASP-52 to be 16.4 days.

Spot crossing events have been modelled with a variety of techniques, such as \textsc{prism} \citep{Reed2013}, \textsc{soap-t} \citep{Oshagh2013} and \textsc{spotrod} \citep{Beky2014}. For this analysis we used \textsc{spotrod} to generate the spot affected light curves and wrote an MCMC wrapper around these generated light curves to fit them simultaneously across the bands and with the same long time-scale trends as before. We chose to use \textsc{spotrod} due to the speed of its integration, which uses polar coordinates in the projection plane. The integration with respect to the polar coordinate is done analytically so that only the integration with respect to the radial coordinate needs to be performed numerically. To calculate the projection of the planet on the stellar surface, \textsc{spotrod} calculates the arrays of planar orbital elements $\xi$ and $\eta$, using the formalism of \cite{Pal2009}, and assumes the same limb darkening law for the spot as for the star.

The system parameters were again fit across the three light curves simultaneously (although this time fitting for impact parameter, $b$, rather than the inclination, as required by \textsc{spotrod}) and with the addition of the parameters defining the spots. The fitting of one spot was tested but was unable to fit both bumps on either side of the transit mid-point, therefore two spots were used in further analysis. The parameters defining each spot were the longitude, latitude, radius ratio of spot to star, and ratio of the spot flux to stellar flux (with 1 being a spot with the same flux as the pristine photosphere and 0 being a spot with zero flux). We held $u1$ fixed as before but now also put Gaussian priors on $u2$ with means equal to the values from \cite{Claret2011} and standard deviations from the propagated errors in the effective temperature and surface gravity of the host star. This prior was necessary as the limb darkening and spot models can play off each other in trying to fit the transit shape.

The MCMC was initiated with the system parameters equal to those in \cite{Hebrard2013} and was run for 10000 steps in burn in and another 10000 steps in the production run. There were 31 fitted parameters with 124 walkers. The parameters governing a spot's characteristics are correlated with one another. The correlation between spot contrast and size has been shown previously by e.g. \cite{Pont2007}, \cite{Wolter2009}, \cite{Reed2013} and \cite{Beky2014}. We initiated the MCMC starting positions with spots at various, randomly selected, latitudes on the stellar surface so as not to bias the fits and test for convergence. After the burn in phase the error bars in the data points were rescaled to give a reduced $\chi^2$ of unity.

After the first MCMC chain, a second MCMC was run but this time with the parameters that were tied across channels fixed to the results from the first run ($a/R_{*}$, $b$, $T_{0}$ and the spot sizes and positions). Correlations with these parameters cause $R_{P}/R_{*}$ to move up and down together across the three wavelengths, contributing to the uncertainty in the absolute planetary radius in each of the bands. Since we are concerned with the shape of the transmission spectrum, we are interested in the relative radii between the bands and not the absolute planetary radius, thus motivating the second run of the MCMC with fixed system parameters.

\begin{figure*}
\centering
\includegraphics[scale=0.5]{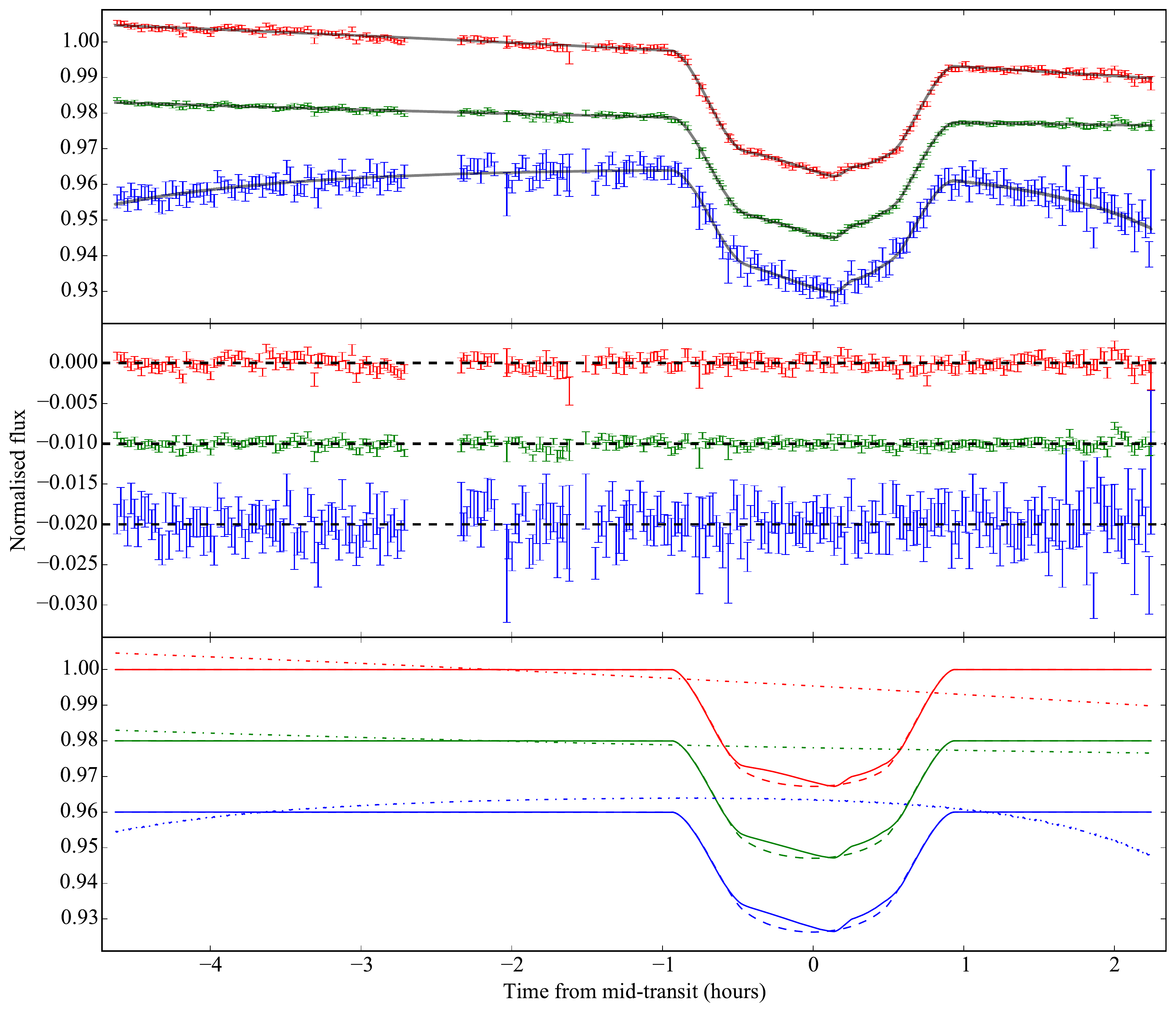}
\caption{MCMC fit of spot model with the occultation of two areas of stellar activity at temperatures lower than the pristine photosphere. The top panel shows the best fitting model for each of the three wavelengths from NaI (top), $g'$ (middle) and $u'$ (bottom), offset by -0.02 and -0.04. The middle panel shows the residuals from these fits, offset by -0.01 and -0.02, whilst the bottom panel shows the constituents of each model, offset by -0.02 and -0.04. The solid line shows the best fitting spot model with the dashed line representing how the model would appear in the absence of spots. The dot-dashed line shows the long time-scale term fitted by a second order polynomial in each of the NaI and $g'$ bands and by a function of airmass in the $u'$ band.}
\label{fig:spots}
\end{figure*}

We present the best fitting spot model in Fig. \ref{fig:spots}, after the second MCMC run, with the results in Table \ref{tab:fit_results} and transmission spectrum in Fig. \ref{fig:trans_spec} (blue squares). 

\begin{figure*}
\centering
\includegraphics[scale=0.4]{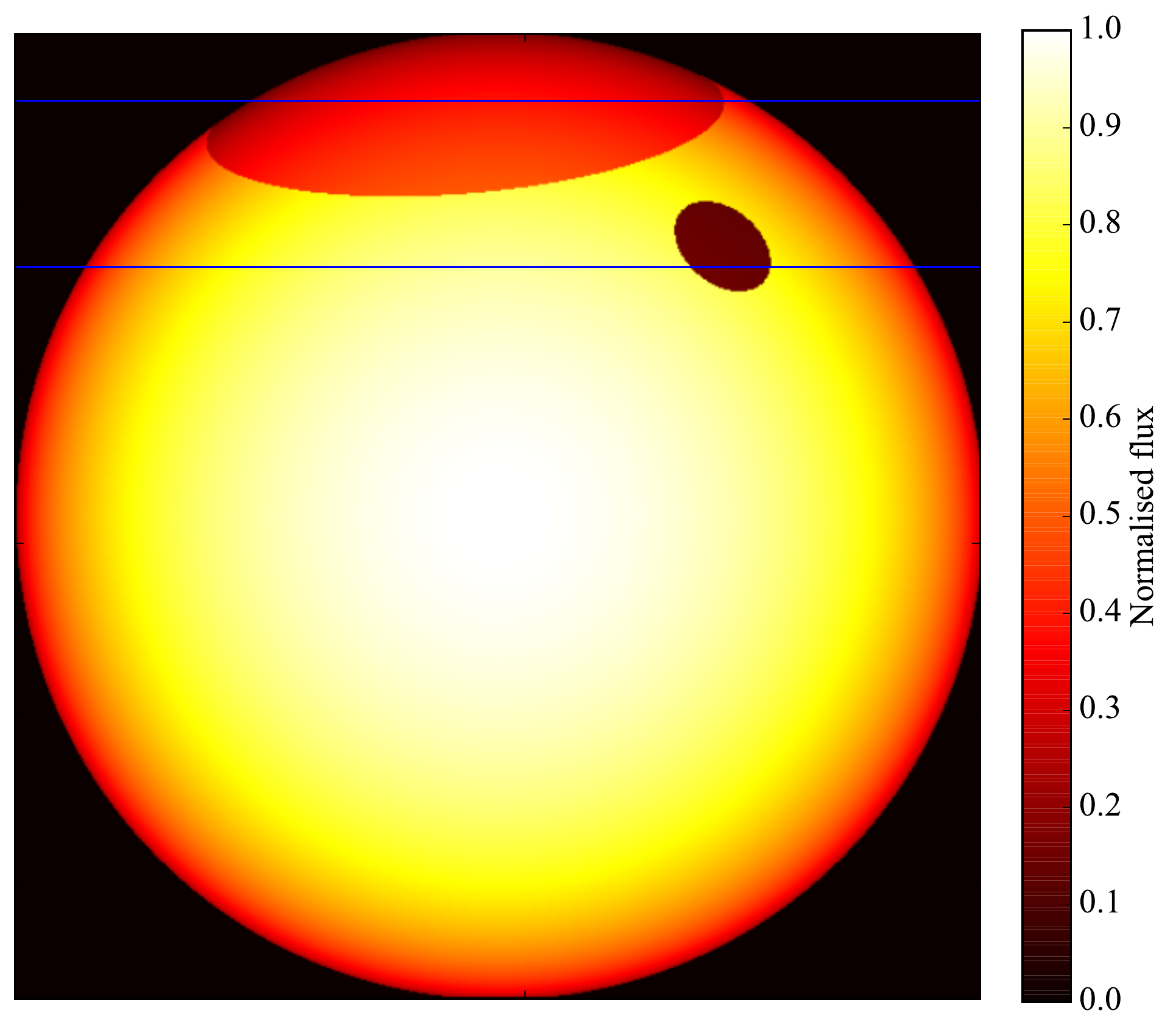}
\includegraphics[scale=0.38]{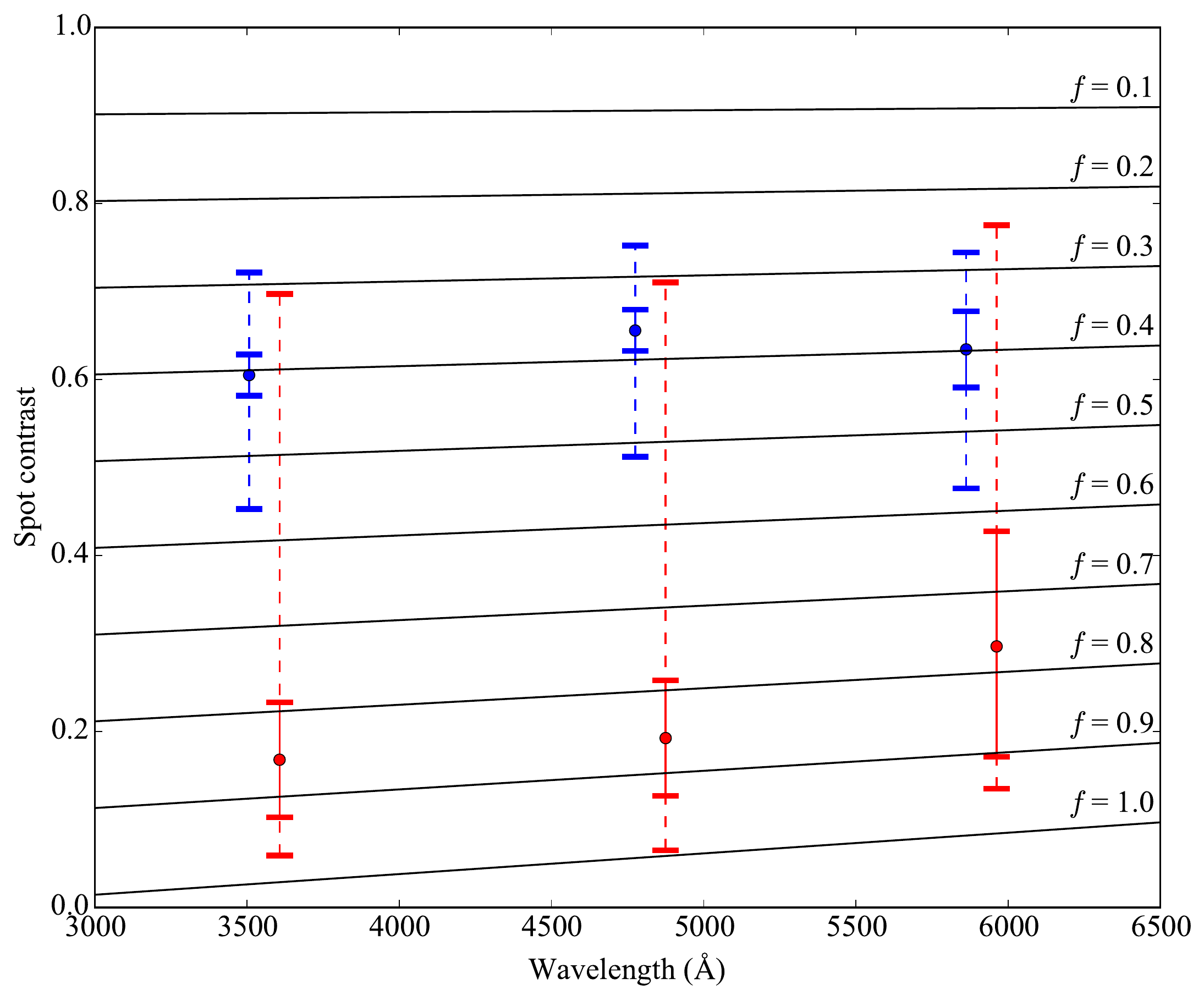}
\caption{\textbf{Left panel:} Schematic illustration of how the stellar surface may appear with the active regions derived from the fitting of the spot model and contrasts consistent with those in the $g'$ band. The blue lines indicate the planet's transit chord. \textbf{Right panel:} Plot of the derived spot contrasts for both spot \#1 (blue) and spot \#2 (red) in each of the three wavelengths, offset in the x-axis by -50 and +50\AA ~respectively for clarity. The dashed error bars show the error in the absolute spot contrasts whilst the solid error bars give the relative uncertainties between the bands for our best fitting system parameters. Over plotted are the calculated contrasts for a spot with $\Delta$T = 1500\,K (consistent with predictions for K stars) with varying filling factors $f$ (solid black lines).}
\label{fig:stellar_surface}
\end{figure*}

With the sizes and contrasts calculated from \textsc{spotrod}, we were able to create a schematic of the stellar surface (Fig. \ref{fig:stellar_surface}, left panel) and consider what filling factor would reproduce the derived contrasts (section \ref{sec:spot_properties} and Fig. \ref{fig:stellar_surface}, right panel). This figure displays the large regions of stellar activity along the transit chord. The second spot crossing event comprises of a smaller region of higher contrast (0.2 in the $g'$ band, Table \ref{tab:fit_results}). The error in the contrast of this dark spot also allowed for a larger, less dark spot.

\subsection{Fitting of bright region model}
\label{sec:facula_model}

As an alternative interpretation, the in-transit anomaly was also modelled as a bright feature analogous to Solar faculae. Solar faculae are bright regions on the solar photosphere which display limb brightening behaviour (e.g. \citealt{Unruh1999}). They are often co-spatial with sunspots but not perfectly so (e.g. \citealt{Haywood2016}). 

The effects of occultations of bright regions in exoplanet transits have been discussed by \cite{Oshagh2014} and could lead to an observable anomaly in transit data. There has not yet been any conclusive evidence of a facula occultation in a transit light curve although \cite{Mohler2013} found evidence for a hot spot in GROND photometry of HATS-2b. They detected a bright feature in the Sloan-$g$ band, which covered the Ca\,II lines, that was consistent with a chromospheric plage occultation.

\begin{figure*}
\centering
\includegraphics[scale=0.5]{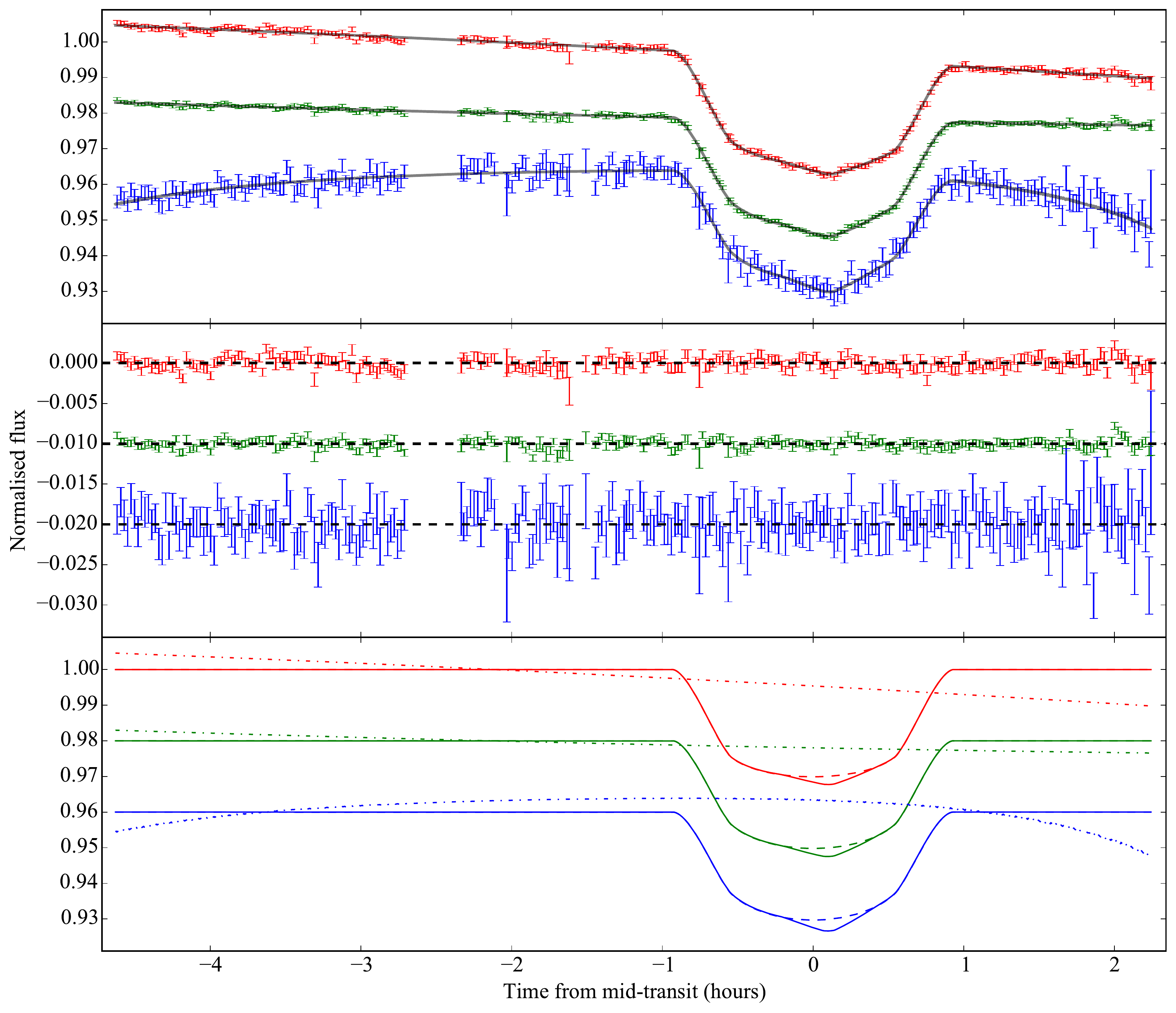}
\caption{The same representation as in Fig. \protect\ref{fig:spots} but this time showing the result of the fitting of a bright spot, analogous to a facula, occulted on the stellar surface. This bright spot leads to a dip in the in-transit data (bottom panel, solid lines) as compared to a transit free of such spots (bottom panel, dashed line).}
\label{fig:plages}
\end{figure*}

\textsc{spotrod} was also used to model the facula scenario but instead of two individual spots with flux ratios $<1$, a single feature was modelled with a flux ratio of $>1$. This fitting method produced the fits seen in Fig. \ref{fig:plages} with the results in Table \ref{tab:fit_results}. This model was able to reproduce the in-transit anomaly with an equally good fit as the two spot model but with six fewer parameters. The transmission spectrum resulting from the facula model is also shown in Fig. \ref{fig:trans_spec} (red triangles). In contrast to the flat transmission spectrum resulting from the fitting of spots, the fitting of a facula led to a slope in the planetary radius increasing towards the red (Fig. \ref{fig:trans_spec}).

\begin{table*}
\caption{Results from fitting the transit light curves with both spots and faculae to model the stellar activity. Note: $x$ position of the spots can take values between $\pm1$ which are analogous to longitudes of $\pm90^{\circ}$ whilst $y$ position takes values between 1 and 0 corresponding to latitudes of $\pm90^{\circ}$. Spot contrast is defined as the ratio of spot flux to stellar flux so a value of unity gives a spot of equal brightness to the surrounding photosphere.}
\label{tab:fit_results}
\centering
\begin{tabular}{l c c c c }
\hline \hline
Parameter 													& Symbol 			& $u'$ & $g'$ & NaI \\

\hline \hline
\emph{2 spot model} & & & &  \\ \hline
Time of mid-transit (MJD)$^a$ & $T_0$ & --- & 56178.12742 $ \pm 0.00005$ & --- \\
Scaled semi major axis$^a$ & $a/R_{*}$ & --- & 6.99 $ \pm 0.04$ & --- \\
Impact parameter$^a$ & $b$ & ---  & 0.656 $^{+0.006}_{-0.007}$ &	--- \\
Radius ratio & $R_{P}/R_{*}$ & 0.1759 $^{+0.0005}_{-0.0004}$ & 0.1751 $ \pm 0.0004$ & 0.1757 $ \pm 0.0008$ \\
Limb darkening coeff.$^b$ & $u_1$ & 0.643 & 0.529 & 0.381 \\
Limb darkening coeff. & $u_2$ & 0.169 $ \pm 0.008$ & 0.233 $ \pm 0.008$ & 0.290 $^{+0.009}_{-0.010}$ \\
Spot 1 x position$^a$ & $x_{1}$ & --- & -0.19 $ \pm 0.03$ & --- \\
Spot 1 y position$^a$ & $y_{1}$ & --- & 0.91 $^{+0.03}_{-0.04}$ &	--- \\
Spot 1 radius ratio$^a$ & $r_{1}/R_{*}$	& ---  &	0.36 $^{+0.04}_{-0.03}$ & --- \\
Spot 1 contrast & $\rho_{1}$ & 0.61 $ \pm 0.02$ & 0.66 $ \pm 0.02$  & 0.63 $ \pm 0.04$ \\
Spot 2 x position$^a$ & $x_{2}$ & ---  & 0.38 $^{+0.02}_{-0.007}$ &	---	\\
Spot 2 y position$^a$ & $y_{2}$ & --- & 0.69 $^{+0.02}_{-0.03}$ &	---	\\
Spot 2 radius ratio$^a$ & $r_{2}/R_{*}$	& --- &  0.07 $^{+0.08}_{-0.006}$	& --- \\
Spot 2 contrast & $\rho_{2}$ & 0.17 $\pm 0.07$ & 0.19 $\pm 0.07$ & 0.3 $\pm 0.1$ \\

\hline \hline
\emph{Facula model} & & & &  \\ \hline
Time of mid-transit (MJD)$^a$ & $T_0$ & --- & 56178.12740 $ \pm 0.00004$ & --- \\
Scaled semi major axis$^a$ & $a/R_{*}$ & --- & 7.23 $ \pm 0.05$ & ---	\\
Impact parameter$^a$ & $b$ & --- & 0.593 $^{+0.008}_{-0.009}$ &	---	\\
Radius ratio & $R_{P}/R_{*}$ 	& 0.1632 $\pm 0.0003$ & $0.1641 \pm 0.0003 $& 0.1657 $\pm 0.0006$ \\
Limb darkening coeff.$^b$ & $u_1$ & 0.643 & 0.529 & 0.381 \\
Limb darkening coeff. & $u_2$ & 0.161 $\pm 0.009$ & 0.239 $ \pm 0.009$ &  $0.29 \pm 0.01$ \\
Facula x position$^a$ & $x_{1}$ & --- &	0.123 $^{+0.007}_{-0.008}$		&	---		\\
Facula y position$^a$ & $y_{1}$ & --- &	0.62 $ \pm 0.03$		&	---			\\
Facula radius ratio$^a$ & $r/R_{*}$ &  --- &	0.20 $^{+0.01}_{-0.02}$			&	---	\\
Facula contrast & $\rho$ & 1.119 $\pm 0.007$ & 1.089 $\pm 0.008$ & $1.08 \pm 0.01$ \\

\hline
\hline\\

\multicolumn{5}{l}{$^a$ parameter is shared and fit across channels with the result quoted in the $g'$ column only and is held fixed at this}\\
\multicolumn{5}{l}{  value for the second MCMC run. The errors on all other parameters are those after the second MCMC run.} \\
\multicolumn{5}{l}{$^b$ parameter not fit by model.} \\
\multicolumn{5}{l}{$^c$ is the combined $\chi^2$ across the three wavelengths after the first MCMC run and before rescaling of the error bars,}\\
\multicolumn{5}{l}{  with d.o.f. giving the degrees of freedom (= number of data points - number of parameters - 1).}\\

\end{tabular}
 
\end{table*}

\section{Discussion}

\subsection{Spots or faculae?}
\label{sec:spots_vs_faculae}

It is difficult to distinguish between the spot and facula models of sections \ref{sec:spot_model} and \ref{sec:facula_model} using the quality of the fits alone. Application of the Bayesian Information Criterion (BIC; given by $BIC = \chi^2 + k\ln N$ with $k$ free parameters and $N$ data points) favours the facula model because of the similar $\chi^2$ and 6 fewer free parameters, with a value of $> 10$ lower than that of the spot model. However, our prior knowledge of transit light curves tends to favour the spot model because spots are more commonly detected (e.g. e.g. \citealt{Pont2007}; \citealt{Sing2011}; \citealt{Mancini2013}; \citealt{Reed2013}) than faculae or bright regions (e.g. \citealt{Mohler2013}).

A second approach to distinguishing between these models is to compare our fitted planet radius with that from independent studies. It can be seen in the model light curves of Fig. \ref{fig:spots}
 and \ref{fig:plages} that our spot and facula models imply different underlying depths of transit. In the spot model the anomaly is treated as two bumps (bottom panel of Fig. \ref{fig:spots}), and so the underlying transit is deeper than in the facula model, where the anomaly is treated as a single dip (Fig. \ref{fig:plages}). This difference can be seen in our fitted planet/star radius ratios (Table \ref{tab:fit_results}) where the spot model implies a planet radius that is 15 sigma larger than the value from the facula model. The planet/star radius ratio has been previously measured by \cite{Hebrard2013} using 6 individual transits, and found to be $0.1646 \pm 0.0012$. This is consistent with our value from the facula model (within 1 sigma), and inconsistent with our value from the spot model ($>6$ sigma discrepancy).

While this paper was under review, another study of WASP-52b was published on arXiv \citep{Mancini_wasp52} including a further 8 transits. Spots were clearly detected in 5 of these light curves and the system parameters measured accounted for these spots. Again, we find that the planet/star radius ratio is consistent with a facula model (within 1 sigma) and inconsistent with our spot model (12 sigma discrepancy).

We also consider the possibility that unocculted spots could be the cause of the discrepant $R_P/R*$ values we derive from the spot model but find this would require a total dimming of the star of 12\%, which we consider unlikely. This is discussed in more detail in section \ref{sec:unocculted}.

We conclude that the comparison of our planet/star radius ratios with independent studies strongly favours the presence of a facula in our observed transit. Nevertheless, in sections \ref{sec:spot_properties} and \ref{sec:facula_properties} we discuss the implications of both models.

\subsubsection{Spot properties}
\label{sec:spot_properties}

The best fitting spot model indicates two large regions of stellar activity (Fig. \ref{fig:stellar_surface}, left panel) which are consistent with the majority of the in-transit data being activity affected (Fig. \ref{fig:spots}). 

Although spot modelling codes, including \textsc{spotrod}, fit spots as circular areas on the stellar surface, in reality these may be areas of several small spots which may be arranged in more complex configurations on the stellar surface. If the features in our light curves really were single spots with temperatures consistent with those expected for K stars, we would expect to see large amplitude bumps and a strong colour dependence. When considering the spots resulting from our modelling however, we believe that the spots are more accurately interpreted as active regions, with spots and pristine photosphere contained within, instead of two large individual spots (Fig. \ref{fig:stellar_surface}).

When considering these complexes of smaller star spots, it is useful to consider the area ratio of spots to uncontaminated photosphere within each of these regions, which we define here as the filling factor, $f$. This quantity can be related to the spot's contrast, $\rho$, through 

\begin{equation}
\label{eq:filling_factor}
\rho = \frac{f F_{\bullet}(\lambda) +  (1-f)F_{*}(\lambda)}{F_{*}(\lambda)}
\end{equation}

\noindent
where $F_{\bullet}$ is the flux of the spot and $F_{*}$ is the flux of the star.

\cite{Berdyugina2005} plotted the observed temperature differences of spots for several different stellar effective temperatures. For K stars, the spot temperature differences ($\Delta$T) lie in the region of 1250 -- 1500\,K.  

Due to the degeneracy between spot size and contrast the spot contrasts have large error bars in Fig. \ref{fig:stellar_surface} (right panel). However, by fixing the spot sizes and positions at the best fitting values from the first MCMC run, the second run gave the relative uncertainty in the spot contrasts across the bands rather than the absolute uncertainty, as with the planetary radii. When considering the relative spot contrasts (Fig. \ref{fig:stellar_surface}, right panel, solid error bars) it can be seen that the contrasts and colour dependences of these active regions are consistent with a $\Delta$T of 1500\,K, given a single filling factor $f$.

For the smaller spot (Fig. \ref{fig:stellar_surface}, right panel, red error bars) we infer a  higher filling factor (approximately 0.85), and for the larger spot (Fig. \ref{fig:stellar_surface}, right panel, blue error bars) a lower filling factor (approximately 0.4). The larger spot can be understood as a large active region on the stellar surface with a relatively low density of smaller dark spots. The smaller spot however has a high density of smaller dark spots leading to the greater contrast.

\subsubsection{Facula properties}
\label{sec:facula_properties}

As discussed in section \ref{sec:spots_vs_faculae} the occultation of a bright region, analogous to a Solar facula, is our favoured interpretation for these data. Unlike spots, the flux contrasts of faculae increase at the stellar limb as they display limb brightening behaviour (e.g. \citealt{Unruh1999}) and so the assumption that the modelled spots follow the same limb darkening as the star does not hold in this case. Therefore Eqn. \ref{eq:filling_factor} cannot be used to calculate the temperature of the facula so we report only the flux contrasts of the facula in Table \ref{tab:fit_results} which indicate the hot spot is $\sim 10\%$ brighter in all three bands. Solar faculae are also able to display such contrasts however, at the Solar limb with a high viewing angle (e.g. \citealt{Unruh1999}; \citealt{Ahern2000}). At the viewing angle of our facula, we would only expect to see a $\sim3\%$ contrast if this were a Solar facula (e.g. \citealt{Ahern2000}). However, since WASP-52 is a more active and later type star it is possible it could display higher contrast faculae.

\subsection{Measuring the residual red noise}

\begin{figure}
\centering
\includegraphics[scale=0.38]{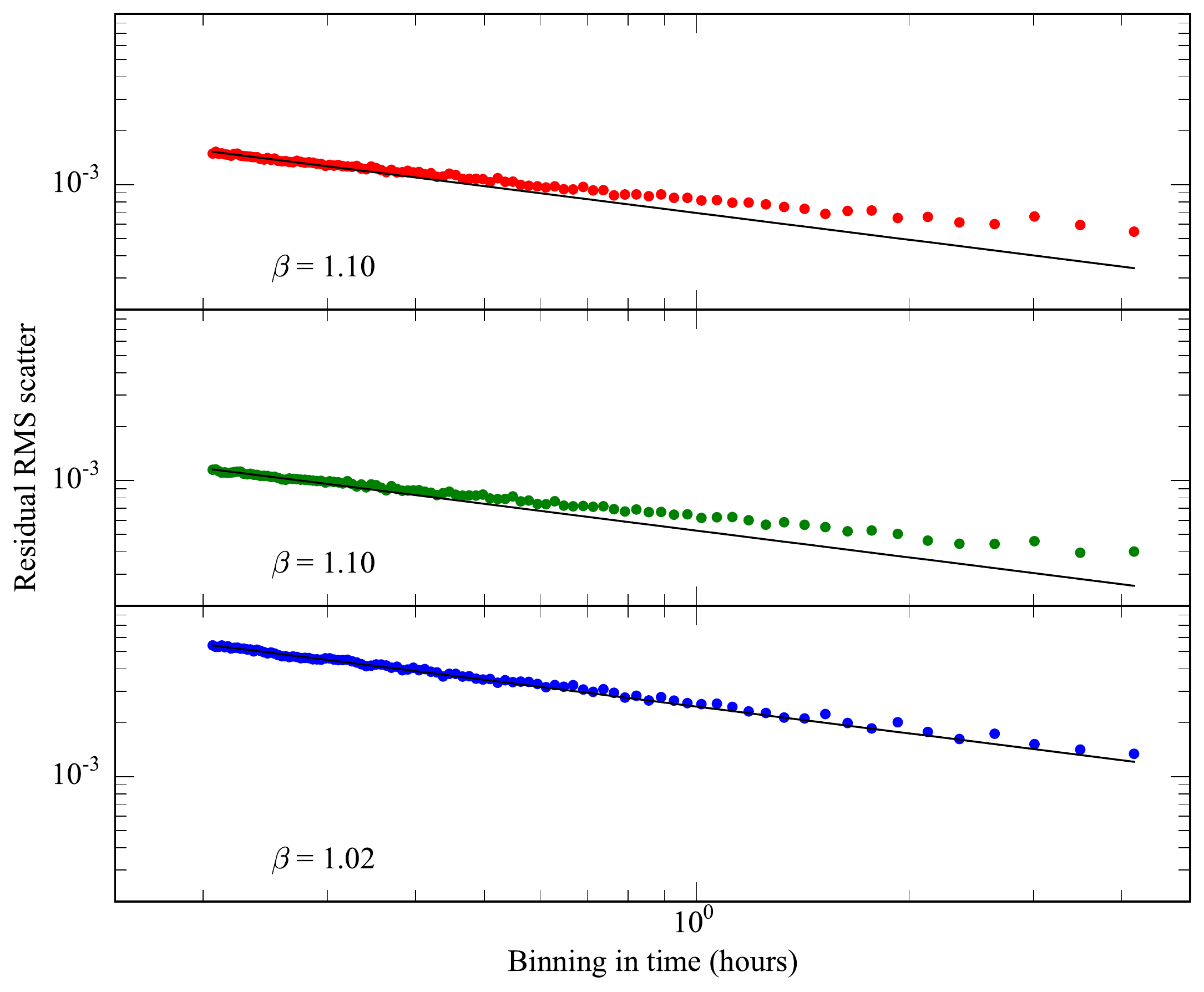}
\caption{The residual RMS scatter once the spot model has been subtracted for NaI (top panel), $g'$ (middle panel) and $u'$ (bottom panel). The error bars on the data points were rescaled to give a $\chi^{2}_{\nu}$ = 1. $\beta$ quantifies how the measured noise deviates from pure Gaussian white noise (solid black lines).}
\label{fig:rms_vs_bins}
\end{figure}

In order to measure the systematic red noise in the residuals, we observed how the fractional root mean square (RMS) of the residuals with respect to the flux varied with binning of the data. Fig. \ref{fig:rms_vs_bins} shows the RMS for a number of different bin sizes. The line overlaid corresponds to pure, Gaussian, white noise with a gradient equal to $1/\sqrt{N}$ where $N$ is the number of points per bin. The factor, $\beta$, quantifies how the actual binning gradient deviates from pure white noise \citep{Winn2008}. The $\beta$ factors found here show that the residuals do deviate from pure white noise but with amplitudes of only 0.5 mmag.

\subsection{Unocculted spots}
\label{sec:unocculted}

Just as the effects of occulted spots or bright regions must be taken into account when calculating the planetary radius, so must the effects of those regions which are not seen in the transit. As the photometric modulation of Sun-like stars with high magnetic activity is dominated by dark spots (e.g. \citealt{Shapiro2014}), we only consider the effect of unocculted spots on our derived transit depths. Unocculted spots cause the observed stellar flux to decrease, increasing the transit depth as compared to a spot free surface. This effect needs to be taken into account with a wavelength dependent depth correction.

Following the formalism of \cite{Sing2011}, the variation in transit depth, $\Delta d/d$, can be related to the fractional decrease of the stellar flux due to unocculted spots at a reference wavelength $\lambda_0$, $\Delta f(\lambda_{0},t)$, the flux of the pristine stellar disc, $F_{\lambda}^{T_{star}}$, and the flux of the spot, $F_{\lambda}^{T_{spot}}$, through

\begin{equation}
\frac{\Delta d}{d} = \Delta f(\lambda_{0},t) \left(1 - \frac{F_{\lambda}^{T_{spot}}}{F_{\lambda}^{T_{star}}}\right)/\left(1 - \frac{F_{\lambda_0}^{T_{spot}}}{F_{\lambda_0}^{T_{star}}}\right)
\end{equation}

\noindent
leading to a variation in the ratio of planet radius to stellar radius of

\begin{equation}
\Delta(R_{p}/R_{star}) \approx \frac{1}{2}\frac{\Delta d}{d}(R_p/R_{star})
\end{equation}

We used stellar atmosphere models (ATLAS9, \citealt{Kurucz1993}) to generate synthetic spectra of the star and a spot with the maximum temperature contrast of 1500\,K cooler than the surrounding photosphere \citep{Berdyugina2005}.

If we rearrange the above equations for $\Delta f(\lambda_{0},t)$, we can estimate the percentage of total dimming required to bring our derived $R_P/R_*$ values into agreement with those of \cite{Hebrard2013}. We find that the total dimming must be $>12\%$ at a reference wavelength of 6000\,\AA, considerably higher than the 1\% amplitude measured by \cite{Hebrard2013}. As a result of this, we do not believe that unocculted spots are the cause of the discrepancy in the $R_P/R_*$ values resulting from the spot model.

To calculate what effect unocculted spots have on the shape of our transmission spectra, we used a total dimming of 1\% at a reference wavelength of 6000\,\AA \,to calculate $\Delta(R_{P}/R_{*})$ in each of our 3 light curves. The difference in $\Delta(R_{P}/R_{*})$ between the red and the blue light curves is 10 times smaller than the 1 sigma error bars and so unocculted spots do not affect the shape of our transmission spectrum (Fig. \ref{fig:trans_spec}) or the conclusions we draw from it.

\subsection{Transmission spectrum}
\label{sec:trans_spec}

\begin{figure*}
\centering
\includegraphics[scale=0.6]{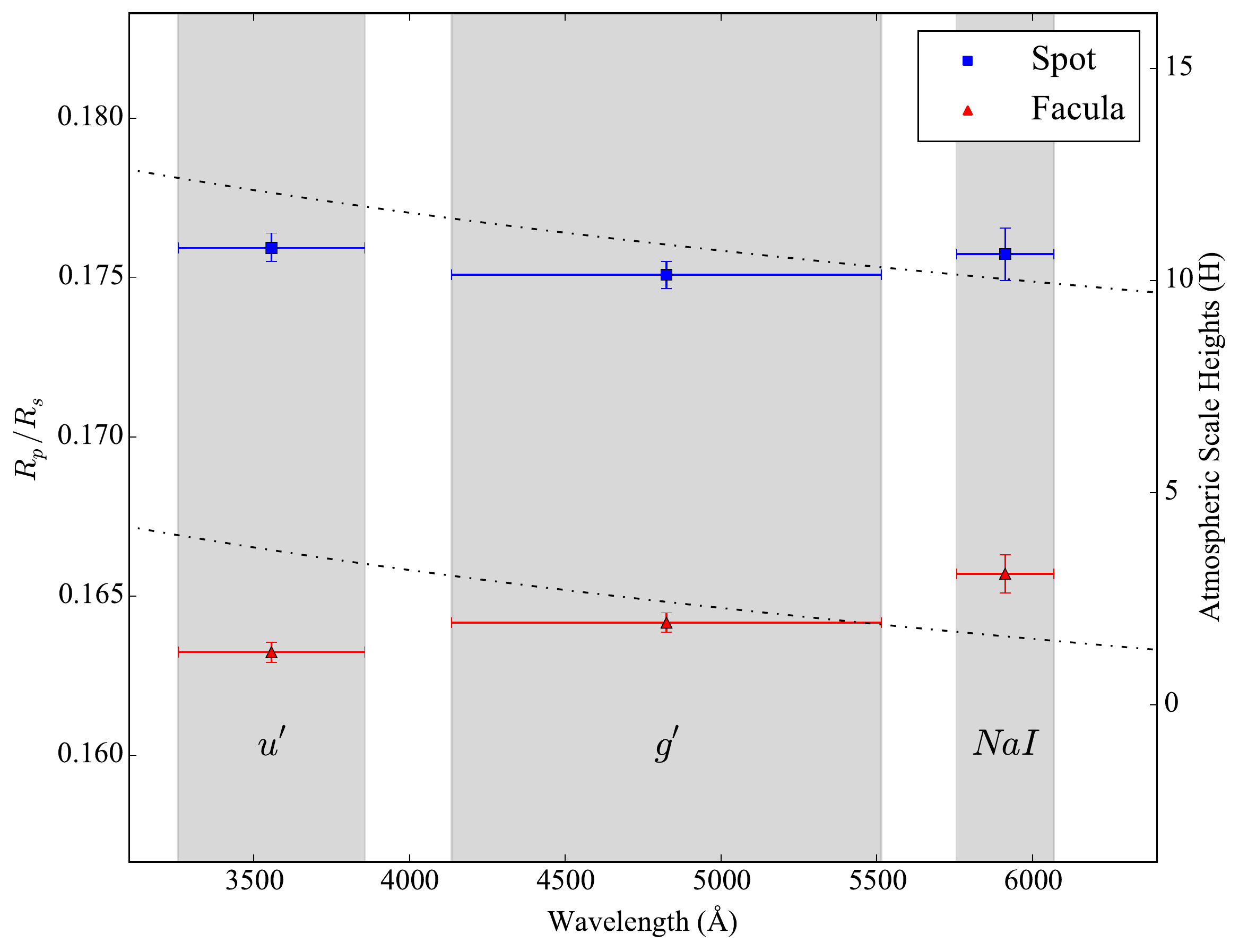}
\caption{Transmission spectrum of WASP-52b. The errors in the x axis are quoted as the full width half maxima of the filters used; $u'$, $g'$ and NaI. The results from both the spot (blue squares) and facula (red triangles) models are included. Over-plotted is a Rayleigh scattering slope (black dot-dashed line) which is ruled out with $> 3\sigma$ significance. The scale height axis is calculated using the value for WASP-52b's surface gravity from \protect\cite{Hebrard2013}.}
\label{fig:trans_spec}
\end{figure*}

Figure \ref{fig:trans_spec} displays the derived transmission spectrum for WASP-52b. On this plot are the results from modelling the in-transit anomaly both as spots and as a single facula. The errors in the transmission spectra are those after fitting with the wavelength independent parameters held fixed at the best fitting values from the first MCMC run ($b$, $a/R_{*}$, $T_0$, spot/facula sizes and positions). This resulted in errors of less than one planetary atmospheric scale height.

In the presence of Rayleigh scattering, the expected slope of the planetary radius as a function of wavelength is given by \citep{Etangs2008_HD189}:

\begin{equation}
\frac{dR_{p}}{d\ln\lambda} = \frac{k}{\mu g} \alpha T
\end{equation}
\noindent
where $\mu$ is the mean molecular mass of an atmospheric particle taken to be 2.3 times the mass of a proton, $k$ is the Boltzmann constant, $g$ is the planet's surface gravity, $\alpha = -4$ as expected for Rayleigh scattering, and $T$ we take as the equilibrium temperature.

Using our derived values for the planet's surface gravity (Table \ref{tab:wasp52}) we are able to rule out Rayleigh scattering in this atmosphere with $> 3\sigma$ confidence (Fig. \ref{fig:trans_spec}). The spot model gives a $\chi^2$ of 14.52 for Rayleigh scattering and 2.00 for a flat line, each with 2 degrees of freedom. Therefore the spot model strongly favours a flat transmission spectrum. For the case of the facula model, the transmission spectrum shows an increase in planetary radius towards the red which is not well fit by either a flat line or Rayleigh scattering. The implied increased radius in the red could indicate the detection of the broad wings of NaI. 

An absence of Rayleigh scattering has been seen in a handful of planets to date, such as HAT-P-32b \citep{Gibson2013_hat32}, WASP-29b \citep{Gibson2013_wasp29} and HAT-P-19b \citep{Mallonn2015}. The absence of Rayleigh scattering and any broad spectral features has been attributed to clouds in the upper atmosphere of the planet obscuring any such features. It could be that clouds are masking the spectral features of WASP-52b also.

\subsection{Updated system parameters}

As outlined in section \ref{sec:spots_vs_faculae}, we compared our results to those of \cite{Hebrard2013} (Table \ref{tab:wasp52}). We note that \cite{Baluev2015} also studied WASP-52b in looking for transit timing variations but the results of the parameters that they fit for, $R_{P}/R_{*}$ and $b$, are within the 1$\sigma$ errors of \cite{Hebrard2013}.

All the parameters in Table \ref{tab:wasp52} marked with $^b$ were derived from the $g'$ band transit light curve alone, following the prescription of \cite{Seager2003}. The planet surface gravity was calculated using the equation from \cite{Southworth2007} which can be derived from the transit light curve but with the addition of the stellar reflex velocity for which we use the value from \cite{Hebrard2013}. The parameters that we have held fixed in our fitting (semi major axis and period) and those which we do not derive are left blank in Table \ref{tab:wasp52}. The parameters marked with $^c$ were derived assuming a stellar radius equal to that of \cite{Hebrard2013}.

As discussed in section \ref{sec:spots_vs_faculae}, the differing implied transit depths of the spot and facula models result in significantly different system parameters (Table \ref{tab:wasp52}). When modelled as two spots, we find a transit depth $>3\sigma$ larger than that of \cite{Hebrard2013}, whereas the facula model results in a transit depth consistent with that of \cite{Hebrard2013}.

\begin{table*}
\caption{Comparison between derived system parameters from this work with WASP-52b's discovery paper \protect\citep{Hebrard2013}.}
\label{tab:wasp52}
\centering
\begin{tabular}{|c|c|c|c|c|c|}
\hline
Parameter (units) & Symbol  & This work (Spots) & This work (Facula) & \cite{Hebrard2013} \\
\hline

Orbital period (days)                    & $P$                     &    --- & ---     & $ 1.7497798  \pm 0.0000012  $          \\           
Semi major axis (AU)                	 & $a$                   &  ---   &  ---        & $ 0.0272     \pm 0.0003     $     \\           
Orbital eccentricity              		 & $e$                                        &  ---  &  ---  &  0 (fixed) \\  
Planetary equilibrium temperature (K) 	 & $T_{\rm P}$             &  --- &  --- & $ 1315       \pm   35       $\\   

Transit epoch (HJD-2450000.0) (days)$^a$     & $T_0$                 &  $6178.62742  \pm 0.00005$    & $6178.62740 \pm 0.00004$    & $ 5793.68143 \pm    0.00009 $        \\           
Scaled stellar radius      $^a$       		 & $R_*/a$                 &  $0.1431 \pm 0.0009$  & $0.1383 \pm 0.0008  $    & $ 0.1355     \pm 0.0020     $         \\           
Impact parameter ($R_*$)   $^a$              & $b$                     &  0.656 $^{+0.006}_{-0.007}$ & $ 0.593 ^{+0.008}_{-0.009}$     & $ 0.60       \pm 0.02       $       \\           
Orbital inclination ($^\circ$)  $^a$         & $i_p$                   &  $84.62 \pm 0.07$    &  $85.30 \pm 0.08 $    & $ 85.35      \pm  0.20      $   \\           

Transit duration (days)$^b$                 & $t_T$                   &  $ 0.0777 \pm 0.0006$ & $0.0772 \pm 0.0006$ & $ 0.0754     \pm 0.0005     $ \\           
Planet/star area ratio$^b$     		 & $(R_{\rm p}/R_*)^2$     &  $0.0306 \pm 0.0005$      & $0.0270 \pm 0.0002 $   & $ 0.0271     \pm 0.0004     $            \\           
Stellar density ($\rho_\odot$)$^b$           & $\rho_*$                &  $1.50 \pm 0.03$  & $ 1.66 \pm 0.03   $   & $ 1.76       \pm 0.08       $ \\           

Stellar radius (R$_\odot$)          & $R_*$                   &   ---  &   ---   & $ 0.79       \pm 0.02       $     \\            

Planet radius (R$_{\rm J}$)$^c$         & $R_{\rm p}$             & $ 1.34 \pm 0.04$ & $ 1.26 \pm 0.03 $ & $ 1.27       \pm 0.03       $\\           

Stellar reflex velocity (km s$^{-1}$)   & $K_1$                  & --- &  --- & $ 0.0843     \pm 0.0030     $  \\ 
Planet surface gravity (cgs)            & $\log g_{\rm p}$        & $ 2.75\pm0.02 $ &  $2.83 \pm 0.02$  & $ 2.81       \pm 0.03       $\\           
Planet density ($\rho_{\rm J}$)$^c$       & $\rho_{\rm p}$          & $ 0.383 \pm 0.024 $ &  $0.204\pm0.009$  & $ 0.22       \pm 0.02       $\\           
Planet mass (M$_{\rm J}$)$^c$           & $M_{\rm p}$             &  $ 0.38\pm0.02$ & $ 0.41 \pm 0.03 $ & $ 0.46       \pm 0.02       $\\

\hline\\

\multicolumn{1}{l}{$^a$ fitted parameter.}\\
\multicolumn{2}{l}{$^b$ derived from the transit light curve alone.}\\
\multicolumn{5}{l}{$^c$ derived using stellar radius from \protect\cite{Hebrard2013}.} \\

\end{tabular}

\end{table*}

\section{Conclusions}

In this paper we have measured the optical transmission spectrum of the highly inflated hot Jupiter WASP-52b. 

Our multi-wavelength light curves, observed with WHT/ULTRACAM, have demonstrated that 0.5 mmag precision can be achieved with high quality ground-based photometry. This leads to errors in our planetary radii of less than one planetary atmospheric scale height, comparable to recent HST observations (e.g. \citealt{Nikolov2014}; \citealt{Sing2015}; \citealt{Fischer2016}). 

Our transit light curves revealed the presence of regions of stellar activity whose in-transit anomalies were modelled using \textsc{spotrod} \citep{Beky2014}. Although modelled as discrete events, it is likely that the planet in fact transits across latitudes of high stellar activity which may be near continuous across the transit chord.

We find this activity can be most simply modelled as a bright region akin to Solar faculae, which results in system parameters consistent with independent studies. The light curves can also be fit with two regions of dark spots but this requires a planet/star radius ratio inconsistent with these studies. As a result, the occultation of a bright region on the host star is the favoured interpretation of the feature seen in the light curves.

We find that Rayleigh scattering is not the dominant source of opacity within the planetary atmosphere. When modelling the in-transit anomalies as spots, we derive a transmission spectrum consistent with wavelength-independent absorption by clouds and find no evidence for any broad sodium absorption although we cannot rule out the presence of the narrow line core. For our favoured facula model, we find an increasing planetary radius towards the red optical which could be interpreted as evidence for the broad wings of NaI.

This work highlights the need for high precision photometry at multiple wavelengths simultaneously to detect and study the effects of stellar activity, in the form of spots and bright regions, on the derived transit parameters.

\vspace{10mm}

We thank the anonymous referee for their helpful comments which improved the quality of the paper.

J.K. and T.L. are supported by Science and Technology Facilities Council (STFC) studentships. P.W. and T.R.M. are supported by an STFC consolidated grant (ST/L00073). D.J.A. acknowledges funding from the European Union Seventh Framework programme (FP7/2007- 2013) under grant agreement No. 313014 (ETAEARTH). V.S.D, S.P.L and ULTRACAM are supported by an STFC consolidated grant (ST/M001350).

\bibliographystyle{mn2e}
\bibliography{Report_ref}

\end{document}